\documentclass[prd,twocolumn,preprintnumbers,superscriptaddress,nofootinbib,showpacs]{revtex4}
\usepackage[a4paper, hdivide={1.91cm,,1.165cm}, vdivide={1.83cm,,2.6cm}]{geometry}

\usepackage{amstext,amssymb}
\usepackage{amsmath}
\usepackage{graphicx}
\usepackage[hyperfootnotes=false]{hyperref}
\usepackage{xspace}
\usepackage{color}
\usepackage{units}
\usepackage{slashed} 

\def \L {\mathcal{L}} 

\definecolor{niceblue}{rgb}{0.15,0.15,0.6}
\definecolor{nicegreen}{rgb}{0.1,0.5,0.1}
\definecolor{Red}{rgb}{1.,0.,0.}

\definecolor{Green}{rgb}{0.2,.7,0.2}

\usepackage{pifont}
\usepackage{multirow}
\usepackage{longtable}

\begin{document}

\title{Gauge coupling unification in a minimal non-supersymmetric $E_6$ GUT}


\author{Chandini  \surname{Dash}}
\email{dash25chandini@gmail.com}
\affiliation{Department of Physics, Berhampur University, Odisha-760007, India}


\author{Snigdha  \surname{Mishra}}
\email{mishrasnigdha60@gmail.com}
\affiliation{Department of Physics, Berhampur University, Odisha-760007, India}

\author{Sudhanwa \surname{Patra}}
\email{sudhanwa@iitbhilai.ac.in}
\affiliation{Department of Physics, Indian Institute of Technology Bhilai, Raipur-492015, India}

\author{Purushottam \surname{Sahu}}
\email{purushottams@iitbhilai.ac.in}
\affiliation{Department of Physics, Indian Institute of Technology Bhilai, Raipur-492015, India}

\begin{abstract}

We consider a minimal renormalizable non-supersymmetric $E_6$ Grand Unified Theory using fundamental representation $27$ for fermions and scalars. The scalar with adjoint  representation ${78}$ is also taken for direct breaking of $E_{6}$ to SM. The proposed model, guided by TeV-scale vector-like fermions and scalar leptoquark offer successful gauge unification even in the absence of any intermediate symmetry. Embedded with threshold corrections, it is shown to be compatible with the present experimental limit on proton decay lifetime. The unique feature of the model shows that, the GUT threshold corrections to the unification mass, is controlled by superheavy gauge bosons only, thereby minimising the uncertainty of the GUT predictions. The scalar leptoquark and vector-like fermions  residing in $27$ representation can explain flavor physics anomalies like $R_{D^{(\ast)}}$ as reported by the LHCb collaboration and the muon anomalous magnetic moment reported by the recent muon $g-2$ experiment at Fermilab. The model can also predict a sub-eV scale neutrino at one-loop level via exchange of $W$ and $Z$ gauge bosons through MRIS mechanism.

\end{abstract}


\maketitle

\noindent
 {\bf 1. Introduction:}
\label{sec:intro}
Standard Model (SM), a masterpiece woven by Glashow, Weinberg and Salam still remains the most treasured theory of particle physics with the discovery of Higgs boson at LHC. However with some unsolved puzzles associated with it, the Standard Model is unlikely to be the final theory, but an effective theory operative at a high scale. Many of the pressing issues can be adequately addressed in Grand Unified Theories (GUTs), with key predictions on experimental findings on non-zero neutrino masses, dark matter candidature, stability of proton as well as recent flavor physics anomalies etc. $E_6$ GUT \cite{Gursey:1975ki,Shafi:1978gg} is a plausible choice of unified theory of nature, since most of the nice features of the well known GUTs $SU(5)$ \cite{Georgi:1974sy} and $SO(10)$ \cite{Fritzsch:1974nn} are inherent here through its embedding. It has the additional advantage of accommodating some exotic particles contained in the fundamental representation $27_F$, which may circumvent the unsolved issues of particle physics.

 In the present study, we propose a non-supersymmetric $E_6$ GUT in its minimal version, as an unified theory to focus on some of these issues. The novel feature of the present work is that the proposed model, guided by TeV-scale extended SM scenario and GUT threshold corrections, is consistent with proton lifetime bound even in absence of any intermediate symmetry. Unlike the existing GUT models with threshold effects \cite{Babu:2015bna,Schwichtenberg:2018cka,Chakrabortty:2019fov,Dash:2020jlc}, the proposed model shows that, the threshold corrections to the unification mass, is controlled by superheavy gauge bosons only, thereby minimising the uncertainty of the GUT predictions. Further it has the potential to address flavor-physics anomalies, muon $g-2$ anomaly and light neutrino masses. 
 
This paper is organized as follows. In sec-2, the key ingredients of the proposed model has been described taking into account the renormalization group effects. In the next section, we focus on the uncertainty for the unification mass and the GUT coupling due to  threshold effects in tune with the predictions for proton lifetime. The subsequent section is devoted to the phenomenological prediction with reference to the neutrino mass and flavor physics anomalies. The concluding section briefly focuses on the salient features of the model.

\noindent
{\bf 2. Model framework:}
\label{sec:model}
The fundamental representation of
 $E_6$ i.e., the $27_F$ provides unified picture of SM plus additional particles as,
\begin{eqnarray}
27_F&=&Q_L\,(3,2,1/6) + u_R\,(3,1,2/3) + d_R\,(3,1,-1/3)  \nonumber \\
&&+ \ell_L\,(1,2,-1/2) + e_R\,(1,1,-1) +N_R(1,1,0) \nonumber \\
&&+ \mathbb{D}_{L,R}\,(3,1,-1/3) + \Psi_{L,R}\,(1,2,-1/2)+s_L(1,1,0)  \nonumber \\
&=& 15_F (\mbox{SM}) + \mbox{Others}
\end{eqnarray}
where, the numbers inside parenthesis are quantum numbers under SM gauge group $(SU(3)_C, SU(2)_L, U(1)_Y)$. 

\noindent
%
The proposed model contains the simplest symmetry breaking chain as, 
\begin{eqnarray}
E_6 
&&	\stackrel{M_U}{\longrightarrow}SU(3)_C\otimes SU(2)_L\otimes U(1)_{Y} \big(\mbox{SM}\big) \nonumber \\
&&	\stackrel{M_Z}{\longrightarrow} SU(3)_C\otimes U(1)_{Q}
\end{eqnarray}

$E_6$ GUT is spontaneously broken directly to the SM gauge symmetry $\mathbb{G}_{321}$ $(SU(3)_C\otimes SU(2)_L\otimes U(1)_{Y})$ at $M_U$ by SM singlets of ${(27+78)}_{H}$. More explicitly, the breaking is done via the trinification symmetry $\mathbb{G}_{333}$ $(SU(3)_C\otimes SU(3)_L\otimes SU(3)_R)$ channel such that

\begin{eqnarray*}
&&{E_6} \supset SU(3)_C\otimes SU(3)_L\otimes SU(3)_R (\mathbb{G}_{333})\nonumber \\ 
&&\supset SU(3)_C\otimes SU(2)_L\otimes SU(2)_R\otimes U(1)_{Y_{L}}\otimes U(1)_{Y_{R}} (\mathbb{G}_{32211})\nonumber \\
&&\supset SU(3)_C\otimes SU(2)_L\otimes SU(2)_R\otimes U(1)_{B-L} (\mathbb{G}_{3221})\nonumber \\
&&\supset SU(3)_C\otimes SU(2)_L\otimes U(1)_{Y} (\mathbb{G}_{321})
	\label{eq:ModelD}
	\end{eqnarray*}
	
where the $(B-L)$ assignment is given by, $Y_{L}+Y_{R}=B-L$. The subsequent stage of symmetry breaking of SM to low energy theory is done at $M_Z$ scale by assigning non-zero VEV to the known SM Higgs doublet contained in $27_H$. However gauge unification is not possible here like other existing non-SUSY GUT models in the absence of any intermediate symmetry. Therefore to achieve unification, we introduce some exotic particles contained in the fundamental representation $27_{F/H}$ i.e., one copy of vector-like fermions $\mathbb{D}_{L,R}(3,1,-1/3)$, three copies of $\Psi_{L,R}(1,2,-1/2)$ and three copies of sterile fermions $N_{R}(1,1,0)$, $s_{L}(1,1,0)$ as well as scalar $S_{1}(\overline{3},1,1/3)$ along with another conventional doublet $\phi(1,2,-1/2)$, which are operative at a mass scale $M_I$. It is noteworthy to mention that, to maintain minimality, we use only those scalars and fermions which are contained in the fundamental and {adjoint} representations of $E_6$ GUT (i.e., $27_{F/H}$ and $78_{H}$) to reproduce all known SM observables plus other unsolved issues of SM. 


Using the standard procedure, we calculate one-loop (${b_i}$) and two-loop (${b_{ij}}$) $\beta$ coefficients \cite{Jones:1981we} from $M_Z-M_I$ and ${b^\prime_i}$ (${b^\prime_{ij}}$) for mass range $M_I-M_U$ with $i,j=3C,2L,Y$, as presented in TABLE-\ref{tab:beta}. 
\begin{table}[t!] 
\caption{\label{tab:beta} $\beta$ coefficients at one-loop and two-loop levels in present model.}
\centering
\begin{tabular}{l l l}
\hline\hline
Mass Range & one-loop level  & two-loop level \\[1mm]
\hline
$M_Z-M_I$    & $(-7, -\frac{19}{6}, \frac{41}{10})$ & $\begin{pmatrix}
-26 & \frac{9}{2} & \frac{11}{10} \\
12 & \frac{35}{6} & \frac{9}{10} \\
\frac{44}{5} & \frac{27}{10} & \frac{199}{50}
\end{pmatrix}$\\
$M_I-M_U$   & $(-\frac{37}{6}, -1, \frac{86}{15})$ & $\begin{pmatrix}
-\frac{29}{3} & \frac{9}{2} & \frac{41}{30} \\
12 & \frac{65}{2} & \frac{21}{10} \\
\frac{164}{15} & \frac{63}{10} & \frac{721}{150}
\end{pmatrix}$\\
\hline\hline
\end{tabular}
\end{table}



 By solving the renormalization group equations for gauge coupling constants and using the experimental values for $\alpha_{em}$, $\alpha_{s}$ and electroweak mixing angle $\sin^2\theta_W$, the  mass scales $M_I$, $M_U$ and inverse GUT coupling constant $\alpha^{-1}_G$ are calculated.
 It is observed that the one-loop and two-loop values for $M_U$ and $\alpha^{-1}_G$ are very close to each other. Thus, we will consider all our analysis further with one-loop effects only. The estimated values are given as:
$$M_I= 10^{4.23}\,\,\mbox{GeV}, M_U= 10^{13.70}\,\,\mbox{GeV and} \quad \alpha^{-1}_G=35.6985.$$ 
 
 However the unification mass scale is found to be much below the desired value for a viable phenomenology like proton decay lifetime etc.
Thus in order to comply with the present data, we consider the GUT threshold effects at $M_U$, arising from superheavy particles (scalars, fermions and gauge bosons) whose masses differ from the symmetry breaking scale $M_{U}$.


\noindent
{\bf 3. GUT Threshold corrections and estimation of Proton Lifetime:}
%

It is well known that threshold effects give corrections to gauge couplings at the symmetry breaking scale and thereby modify the unification scale and other phenomenological parameters. In the present framework, the 
modifications mainly arise due to the matching condition \cite{Hall:1980kf} at the symmetry breaking scale $M_{U}$, given by:

\begin{eqnarray}
 \alpha^{-1}_{i} (M_{U})&&=\alpha^{-1}_{G} (M_{U})-\frac{{\lambda_{i}^{U}}}{12\pi}
\end{eqnarray}
where $\alpha^{-1}_{G} (M_{U})$ is the inverse GUT coupling constant of the unified group $E_6$, $\alpha^{-1}_{i} (M_{U})$'s are the inverse coupling constants corresponding to $i=3C, 2L, Y$, the Standard Model gauge groups. $\lambda_{i}^{U}$ denote the GUT threshold corrections which arise due to the superheavy scalars with mass $M_H$, vector bosons with mass $M_V$ and fermions with mass $M_F$ around the unification mass scale $M_{U}$, such that,

\begin{eqnarray}
 \lambda_i^{U}&=& 
\mbox{\large Tr} \left[{t_{iV}^2}\right]
  -21\, \mbox{\large Tr} \left[{t_{iV}^2} \eta _{V} \right] \nonumber \\
  &&\hspace*{-1.3cm}+2k \mbox{\large Tr} \left[{t_{iH}^2} \eta _{H}\right] + 8 \kappa\, \mbox{\large Tr} \left[{t_{iF}^2} \eta _{F}\right]
  \label{eq:threshold}
\end{eqnarray}

where $\eta_{V} \equiv \mbox{ln}\frac{M_V}{M_U}, \eta_{H} \equiv \mbox{ln}\frac{M_H}{M_U}, \eta_{F} \equiv \mbox{ln}\frac{M_{F}}{M_U}$. Here $k=\frac{1}{2}(=1)$ for the real (complex) scalar fields and, $\kappa=\frac{1}{2}(=1)$ for Weyl (Dirac) fermions.\\

These superheavy fields follow the Extended Survival Hypothesis and are given in TABLE-\ref{tab:SM}.

\begin{table}[t!] 
\caption{\label{tab:SM} Superheavy scalar, fermion and vector fields that have masses around $M_{\rm GUT}$ in the ${\mathbb{G}_{321}}$ model.}
\centering
\begin{tabular}{l l l}
\hline\hline
&$E_6$ & $\mathbb{G}_{321}$  \\
\hline
Scalars & ${27}_H$ & $H_1({3},{1}, -1/3)$, $H_2({\overline{3}},{1}, 1/3)$, $H_3({1},{2},1/2)$,\\
& &  $H_4({3},{2},1/6)$, $H_5({\overline{3}},{1},-2/3)$, \\
& & $H_6({1},{1},1)$, ${2} H_7({1},{1},0)$\\ \\
 & ${78}_H$ &$4H_8(1,1,0)$, $H_9(8,1,0)$, $H_{10}(1,3,0)$, $H_{11}(1,2,1/2)$,\\
& &  $H_{12}(1,2,-1/2)$, $2H_{13}(1,1,1)$, $2H_{14}(1,1,-1)$,\\
& &  $2H_{15}(\overline{3},{1},-2/3)$, $H_{16}(\overline{3},{1},1/3)$, $2H_{17}(\overline{3},{2},-1/6)$, \\
& &  $H_{18}(\overline{3},{2},5/6)$, $2H_{19}({3},{1},2/3)$, $H_{20}({3},{1},-1/3)$, \\
& &  $2H_{21}({3},{2},1/6)$, $H_{22}({3},{2},-5/6)$\\
\hline
Fermions & ${27}_F$ & {2} $\mathbb{D}_{L}({3},{1}, -1/3)$, {2} $\mathbb{D}_{R}({3},{1}, -1/3)$\\
\hline
Vectors & ${78_V}$ & ${4} V_1({1},{1}, 0)$, $V_2({1},{2}, 1/2)$, $V_3({1},{2},-1/2)$,\\
& &${2} V_4({1},{1},1)$, ${2} V_5({1},{1},-1)$,  ${2} V_6({\overline{3}},{1}, -2/3)$,\\
& &$V_7({\overline{3}},{1}, 1/3)$, ${2} V_8({\overline{3}},{2},-1/6)$,  $V_9({\overline{3}},{2},5/6)$,\\
& &${2} V_{10}({3},{1},2/3)$,  $V_{11}({3},{1},-1/3)$, ${2}V_{12}({3},{2},1/6)$,\\
& &$V_{13}({3},{2},-5/6)$\\
\hline\hline
\end{tabular}
\end{table}

Now considering the one-loop threshold effects, the RGEs for evolution of gauge coupling constants are modified between intermediate scale $M_I$ (around few TeV) to unification mass scale $M_U$ as
\begin{equation}
\hspace*{-0.4cm}\alpha_i^{-1}(M_I)=\alpha_G^{-1}(M_U)+\frac{{b^\prime_i}}{2\pi}{\ln\left(\frac{M_U}{M_I}\right)} - {\frac{\lambda^U_{i}}{12\pi}}
\label{rge-alphainv-MIMU-TH}
\end{equation}

Now using RG equations and following the standard procedure, the analytic expression for the threshold corrections to the intermediate mass scale $M_{I}$, the unification scale $M_{U}$ and the inverse GUT coupling constant $\alpha_G^{-1}$ are obtained as:
 \begin{eqnarray} 
 &&{\Delta} {\large \ln}\left(\frac{M_I}{M_Z}\right)
 = \frac{1}{2112} \bigg[
 155 {\lambda^{U}_{Y}}- 357 {\lambda^{U}_{2L}}+ 202{\lambda^{U}_{3C}} \bigg] \nonumber \\
 &&{\Delta} {\large \ln}\left(\frac{M_U}{M_Z}\right)
 = \frac{1}{264} \bigg[
 5 {\lambda^{U}_{Y}}- 3 {\lambda^{U}_{2L}}- 2{\lambda^{U}_{3C}} \bigg] \nonumber \\
&&
 \Delta \alpha_G^{-1} 
 =\frac{1}{2304\pi}\Big[205\lambda_{Y}^U-243\lambda_{2L}^U+230\lambda_{3C}^U
\Big]
 \label{eq:LalphaG}
 \end{eqnarray}
where, $\lambda_{3C}^U$, $\lambda_{2L}^U$ and $\lambda_{Y}^U$ are the GUT threshold contributions for $SU(3)_C$, $SU(2)_L$ and $U(1)_Y$ gauge groups respectively.



Now we assume that all the superheavy scalars and fermions have degenerate mass $M_{H}$ and $M_{F}$ respectively. However referring to TABLE-\ref{tab:SM}, the massive gauge bosons attain non-degenerate masses $M_{V_{X}}$ and $M_{V_{Y}}$ corresponding to the leptoquark gauge bosons ${V_{X}} (=V_{9}=V_{13})$ and ${V_{Y}} (=V_{8}=V_{12})$ respectively. Now for simplicity, we take the mass of the other superheavy gauge bosons to be degenerate with ${V_{X}}$ and ${V_{Y}}$, such that $M_{V_{X}}=M_{V_{1,2,3,4,5}}$ and $M_{V_{Y}}=M_{V_{6,7,10,11}}$. Thus, the threshold effects $\lambda^{U}_{i}$'s reduce to 

\begin{align} 
\lambda^{U}_{3C} & = 9 -21 \left(2 \eta_{V_X} + 7 \eta_{V_Y}\right)+ 29 \eta_{H} + 16 \eta_{F} \\
\lambda^{U}_{2L} & = 10 -21 \left(4 \eta_{V_X} + 6 \eta_{V_Y}\right) + 28 \eta_{H}  \\
\lambda^{U}_{Y} & = 12 -21 \left(8 \eta_{V_X} + 4 \eta_{V_Y}\right) + \frac{142}{5} \eta_{H} + \frac{32}{5} \eta_{F} 
 \end{align} 
with $ \eta_{V_X} \equiv \mbox{ln}\frac{M_{V_X}}{M_U}, 
\eta_{V_Y} \equiv \mbox{ln}\frac{M_{V_Y}}{M_U}$.
The resulting GUT threshold effects for $M_{I}$, $M_{U}$ and $\alpha_G^{-1}$ are obtained as,

\begin{eqnarray}
 &&\hspace*{-0.3cm} {\Delta}{\large \ln}\left(\frac{M_I}{M_Z}\right)=\frac{1}{176}\Big[9+22\eta_H+352\eta_F\nonumber\\
 &&\hspace*{+2.5cm}-378\eta_{V_{X}}+189\eta_{V_{Y}} \Big]
  \nonumber \\
 &&\hspace*{-0.3cm} {\Delta}{\large \ln}\left(\frac{M_U}{M_Z}\right)=\frac{1}{22}\Big[1-42 \eta_{V_{X}}+21 \eta_{V_{Y}} \Big]
 \label{rel:GUT-TH} \nonumber \\
&& \hspace*{-0.3cm}
 \Delta \alpha_G^{-1} 
 =\frac{1}{192\pi}\Big[175+474\eta_H+416\eta_F\nonumber\\
 &&\hspace*{+2.5cm}-1974\eta_{V_{X}}-1701\eta_{V_{Y}} \Big] 
\end{eqnarray}


It is observed that, there is an effective cancellation between superheavy scalars and fermions, such that the unification mass due to GUT threshold is solely controlled by superheavy gauge bosons only.
 However, the inverse GUT coupling constant does depend on the mass of all particles. Now we fine tune the mass parameters $M_{H}$, $M_{F}$, $M_{V_{X}}$ and $M_{V_{Y}}$, so as to obtain the GUT threshold corrected intermediate and unification mass scales, the invese GUT coupling constants and proton lifetime, with the check point that it is compatible with the experimentally accessible proton decay. Here we limit ourselves to the proton decay channel $p \to e^+ \pi^0$, mostly by the exchange of leptoquark gauge bosons. The general expression for proton lifetime can be expressed as, 
\begin{eqnarray}  
\tau_{p\rightarrow e^+ \pi^0} &=&
 \frac{4}{\pi}  \left(\frac{ f^2_\pi}{m_p}\right) \left(\frac{M^4_U}{\alpha^2_G} \right)
         \frac{1}{\alpha^2_H {A^2_R} \left(1+ |{V_{ud}}|^2\right)^2}  \,
\label{lifetime-proton}
\end{eqnarray}
 where the notations have their usual meaning. However, ${A_R}$ contains the model dependent factors such that ${A^2_R}\simeq {A^2_{L}} \left({\mathcal{A}^2_{SL}} + {\mathcal{A}^2_{SR}} \right)$. In the present model ${A_L}=1.25$, is the long distance enhancement factor. 
 The short distance renormalization factor ${\mathcal A_{SL(R)}}$--both for left as well as right-handed effective dimension-6 operator-- derived in the presence of all possible intermediate scales and is a model dependent factor as,
\begin{eqnarray}
&&{\mathcal{A}_{SL(R)}}= {\mathcal{A}}^{321}_{SL(R)} \cdot {\mathcal{A}}^{321}_{SL(R)} 
\end{eqnarray}
where,  
\begin{eqnarray}
& &{\mathcal{A}}^{321}_{SL(R)}=
          \left(\frac{\alpha^{-1}_{i} (M_I)}{\alpha^{-1}_{i} (M_{U})} \right)
          ^{ \frac{\gamma_{L(R)i}}{b^{\prime}_i}}, \quad 
\mbox{i=3C, 2L, 1Y} \, ; \nonumber    \\
& &{\mathcal{A}}^{321}_{SL(R)}= \left(\frac{\alpha^{-1}_{i} (M_Z)}{\alpha^{-1}_{i} (M_{I})} \right)
          ^{ \frac{\gamma_{L(R)i}}{b_i}}, \quad 
\mbox{i=3C, 2L, 1Y} \, 
\label{eq:SD}
\end{eqnarray}
Here $\alpha_i=g^2_i/4\pi$ is the fine structure constant for gauge group $\mathbb{G}_i$. Further $\gamma_{L(R)i}$'s are the anomalous dimensions \cite{Dash:2020jlc} given by
\begin{eqnarray*}
&&\mbox{For}\,  \mathbb {G}_{3_{C}2_{L}1_{Y}}\,, \quad\left\{ \begin{array}{ll}
                  \gamma_L (M_Z)= \left(2, \frac{9}{4},\frac{23}{20}\right)\\
                  \gamma_R (M_Z)= \left(2, \frac{9}{4},\frac{11}{20}\right)
                 \end{array}
                 \right. 
\end{eqnarray*}
and ${b_i}=(-7, -19/6, 41/10)$  (${b^\prime_i}=(-37/6,-1,86/15)$) are the one-loop beta coefficients at different stage of RGEs from $M_Z-M_{I}$ ($M_I-M_U$), respectively. ${\mathcal{A}_{SL}}=2.32$, ${\mathcal{A}_{SR}}=2.20$ are the values without threshold effects. The other parameters take the usual values, $f_\pi=139\, \, \mbox{MeV}$ and $m_p=938.3\, \, \mbox{MeV}$, $\alpha_H = 0.012\, \, \mbox{GeV}^3$, $V_{ud}=0.974$. Thus using eqn.(\ref{lifetime-proton}), the corresponding $\tau_{p}$ can be calculated with threshold corrected $M_{U}$ and $\alpha_{G}^{-1}$ to examine the viability of the model.

\begin{figure}[htb!]
	\centering
	\includegraphics[width=0.45\textwidth]{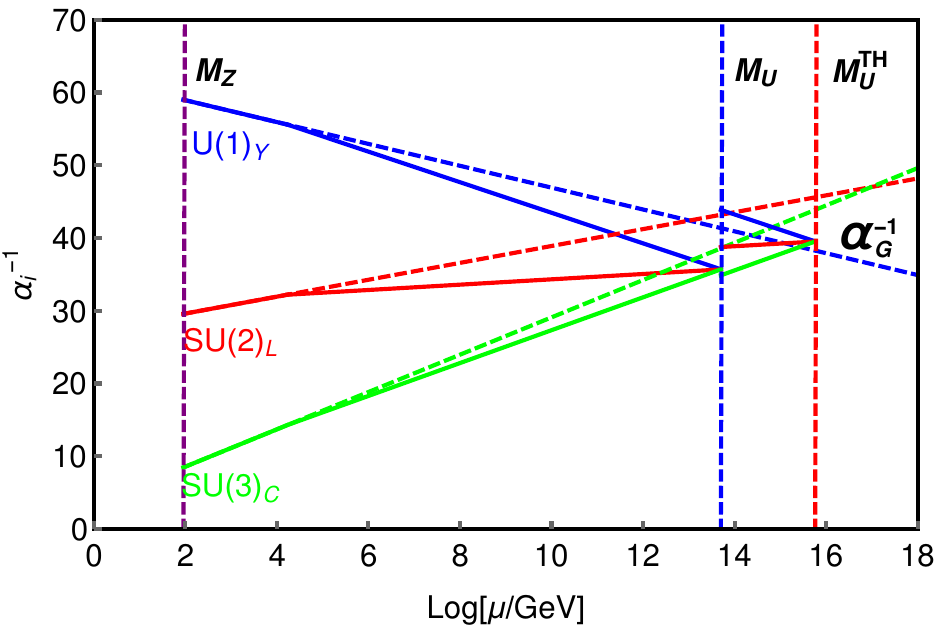}
	\caption{Successful unification of SM gauge couplings within non-SUSY $E_{6}$ GUT with GUT Threshold effects. The unification scale is found to be $M_{U}^{TH}=10^{15.76}$ GeV.}
	\label{fig:unification}
	\end{figure}

 Now we choose the mass parameters $\frac{M_{a}}{M_{U}}$ (with $a=H, F, {V_{X}}, {V_{Y}}$) within the range $[\frac{1}{15},15]$ to calculate $M_I$, $M_U$, $\alpha_G^{-1}$ and $\tau_{p}$. Numerical estimate shows that, with threshold effects we can achieve the unification mass $M_U= 10^{15.59}$ GeV to $10^{15.84}$ GeV, the inverse GUT coupling constant $\alpha_G^{-1}$ within the perturbative range as well as $\tau_{p}$ in comply with present experimentally allowed values \cite{Miura:2016krn,Abe:2011ts,Yokoyama:2017mnt}. The result is shown in TABLE-\ref{tab:pdecay}.
 The corresponding plot is given in FIG.\ref{fig:unification} to show the gauge unification with threshold corrections. Further, to show the variation of proton lifetime $\tau_{p}$ with the mass of the leptoquark gauge bosons, we denote  $\frac{M_{V_{X}}}{M_U}=k_1$ and $\frac{M_{V_{Y}}}{M_U}=k_2$. Thus, we can express the threshold corrections to $M_{U}$ given as,
 
 \begin{eqnarray}
  {\Delta}{\large \ln}\left(\frac{M_U}{M_Z}\right)=\frac{1}{22}\left[1+21\,\, \mbox{ln\,}\left(\frac{k_2}{k^2_1} \right)\right]
 \end{eqnarray}

\begin{figure}[htb!]
\includegraphics[width=0.45\textwidth]{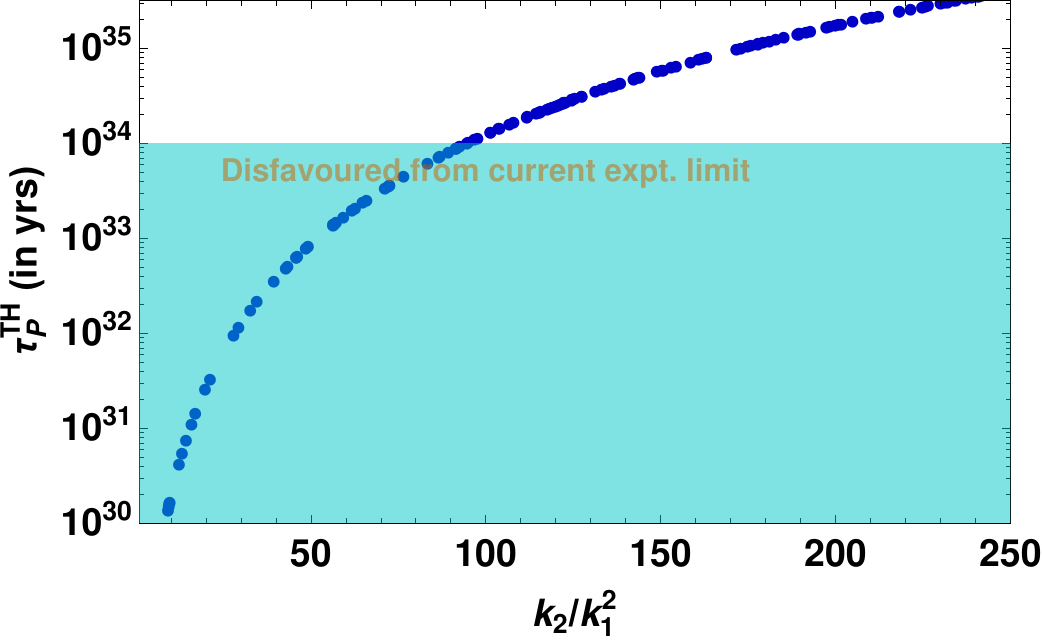}
\caption{Variation of one-loop GUT threshold corrected proton lifetime $\tau_{p}^{TH}$ with the ratio $(\frac{k_{2}}{k_{1}^{2}})$ (the relative mass of the superheavy gauge bosons).}
	\label{fig:taup_plots}
\end{figure}

Now, referring to the TABLE-\ref{tab:pdecay}, for a fixed choice of $M_{H}$ and $M_{F}$, the variation of $\tau_{p}$ with the ratio ($\frac{k_2}{k^2_1}$)(the relative mass of the superheavy gauge bosons) is displayed in the FIG.\ref{fig:taup_plots} to show the favourable region for the proton decay.

 \begin{center}
\begin{table}[t!]
\scriptsize
 \centering
\vspace{-2pt}
\begin{tabular}{||c|c|c|c|c|c|c|c|c||}
\hline \hline
${\frac{M_H}{M_U}}$& ${\frac{M_F}{M_U}}$&${\frac{M_{V_{X}}}{M_U}}$&${\frac{M_{V_{Y}}}{M_U}}$&$\frac{k_{2}}{k_{1}^{2}}$ &  ${M_I}$ & ${M_U}$ & ${\alpha_G^{-1}}$ & ${\tau_{p}}$ \\
 & & $=k_{1}$&$=k_{2}$& &\,\mbox{(GeV)} &\,\mbox{(GeV)} & &$\,\mbox{(years)}$
\\
\hline
$12$ &$\frac{1}{15}$& $\frac{1}{1.7}$&$5.5$&$15.895$&$10^{3.32}$ & $10^{14.87}$ & $33.0028$& $ 2.69 \times 10^{31}$ \\
\hline
$12$ &$\frac{1}{15}$& $\frac{1}{3.0}$&$3.0$&$27$&$10^{3.57}$ & $10^{15.09}$ & $36.5709$& $ 1.57 \times 10^{32}$ \\
\hline
$12$ &$\frac{1}{15}$& $\frac{1}{4.0}$&$4$&$64$&$10^{3.97}$ & $10^{15.45}$ & $36.7011$& $ 4.51\times 10^{33}$ \\
\hline
$12$ &$\frac{1}{15}$& $\frac{1}{4.5}$&$4.5$&$91.125$&$10^{4.14}$ & $10^{15.59}$ & $36.7544$& $ 1.66 \times 10^{34}$ \\

\hline
$12$ &$\frac{1}{15}$&$\frac{1}{5.5}$&$3.1$&$93.775$&$10^{4.15}$ & $10^{15.61}$ & $38.4621$& $ 1.76 \times 10^{34}$ \\
\hline
$12$ &$\frac{1}{15}$&$\frac{1}{6.0}$&$3.3$&$118.8$&$10^{4.26}$ & $10^{15.70}$ & $38.5705$& $4.05 \times 10^{34}$ \\
\hline
$12$ &$\frac{1}{15}$&$\frac{1}{6.5}$&$2.9$&$122.525$&$10^{4.28}$ & $10^{15.72}$ & $39.1969$& $4.66 \times 10^{34}$ \\
\hline
$12$ &$\frac{1}{15}$&$\frac{1}{7.0}$&$2.8$&$137.2$&$10^{4.33}$ & $10^{15.76}$ & $39.5383$& $6.61 \times 10^{34}$ \\
\hline
$12$ &$\frac{1}{15}$&$\frac{1}{7.25}$&$2.9$&$152.431$&$10^{4.38}$ & $10^{15.81}$ & $39.5542$& $ 1.05 \times 10^{35}$ \\
\hline
$12$ &$\frac{1}{15}$&$\frac{1}{10.5}$&$1.5$&$165.375$&$10^{4.42}$ & $10^{15.84}$ & $42.6254$& $1.13 \times 10^{35}$ \\
\hline \hline
\end{tabular}
\caption{Numerically estimated values for $M_I$, $M_U$, $\alpha_G^{-1}$ and $\tau_{p}$ by considering one-loop threshold effects at $M_U$.}
\label{tab:pdecay}
\end{table} 
\end{center}
\vspace*{+0.175cm}
 \noindent
{\bf 4. Low energy phenomenology:\,}
 
In the earlier discussion, we have noted that inclusion of few copies of extra vector like fermions $\mathbb{D}_{L,R} (3,1,-1/3)$, $\Psi_{L,R} (1,2,-1/2)$ and a scalar leptoquark $S_{1}(\overline{3},1,1/3)$ at an intermediate TeV scale, on top of the SM fields, leads to successful gauge coupling unification consistent with proton decay experimental constraints. The relevance of these exotic fermions and scalar can be visualised to explain flavor physics anomalies recently reported by LHCb data and the muon $g-2$ anomaly reported at Fermilab. Further the inclusion of the right-handed neutrinos $N_R$ and extra sterile neutrinos contained in $27_F$ of $E_6$ GUT at intermediate scale or any scale between $M_I$ and $M_U$, without affecting RGEs, can explain sub-eV mass of the light neutrino, consistent with neutrino oscillation data.

 To visualise the generation of neutrino mass, we note that the allowed $E_6$ invariant term $27_F 27_H 27_F$, can only accommodate the Dirac masses, but no Majorana masses for the neutrinos, at tree level. This is obvious due to absence of electroweak scalar triplets or $351^{\prime}$ representation of $E_6$. Thus with minimal $27_{H}$ representation, the relevant mass term contains $m_\text{D} \overline{\nu_L} N_R + M \overline{s_L} N_R$, where $m_\text{D}$ is the Dirac neutrino mass matrix connecting $\nu_L-N_R$, $M$ is the heavy $N_{R}-s_{L}$ mixing matrix. As a result, type-I and type-II contributions are absent in the present framework. However, the Majorana mass terms for $N_R$ and $s_L$ can be generated via radiative mechanism \cite{Cauet:2010ng,Babu:2017xlu}, such that light neutrinos of sub-eV scale, can be obtained via minimal radiative inverse seesaw (MRIS) mechanism \cite{Pilaftsis:1991ug,Dev:2012sg}. The idea is that the physical neutrinos are massless at the tree level while heavy neutrinos $N_R$ and $s_L$ form a heavy Dirac pairs of the order of $M$. In MRIS mechanism, the active neutrino acquires a sub-eV mass at one-loop level via exchange of Standard Model $W$ and $Z$ gauge bosons. 


With reference to recent flavor physics anomalies like lepton flavor universality violation (LFUV) $R_{D^{(\ast)}} = \left. \dfrac{\mathcal{B}(B\to D^{(\ast)} \tau\bar{\nu})}{\mathcal{B}(B\to D^{(\ast)} l \bar{\nu})}\right|_{l\in \{e,\mu\}}$ ~\cite{Fajfer:2012vx}, one needs to invoke a scenario of physics beyond the SM (BSM). In the context of popular BSM scenarios, the leptoquarks (LQs) (scalar or vector) which are colored particles that couple to both quarks and leptons, can be a mediator of new physics (NP) that can accommodate B-physics anomalies\cite{Sakaki:2013bfa,Angelescu:2018tyl,Kirk:2023fin,DeRomeri:2023cjt}. The nice feature of the present model shows that scalar leptoquark ($S_{1}(\overline{3},1,1/3)$)  can explain $R_{D^{(\ast)}}$ which is contained in fundamental representation $27$.


$R_{D^{(\ast)}}$ anomalies has been attributed to specific leptoquark $S_1$ in the literature              \cite{Freytsis:2015qca,Bauer:2015knc,Becirevic:2016oho,Cai:2017wry,Angelescu:2021lln}. It was illustrated in reference \cite{Bauer:2015knc} that a TeV-scale $S_1$ leptoquark is capable of  explaining the $R_{D^{(\ast)}}$ anomalies at tree level.

To fully comprehend the B-physics anomalies, we must first write down the Lagrangian, which includes the interaction terms of the scalar leptoquark $S_{1}(\overline{3},1,1/3)$ and vector-like fermions such as $\mathbb{D}_{L,R} (3,1,-1/3)$, $\Psi_{L,R} (1,2,-1/2)$ with the fermions present in the fundamental representation of $E_{6}$ GUT.
The relevant interaction terms \cite{Dorsner:2016wpm} needed for phenomenological analysis at low energy can therefore be expressed as,

\begin{eqnarray}
  {\L}\supset &&{y_{L}}\overline{Q}{_{L}^{c}}{i}\tau_{2} \ell_LS_{1}+{y^{'}_{L}}\overline{Q}{_{L}^{c}}{i}\tau_{2} \Psi_{L}S_{1}+{y^{''}_{L}}\overline{Q}{_{L}^{c}}{i}\tau_{2} Q_{L}S_{1}^{*}
\nonumber\\
 &&+ {y_{R}}\overline{u}{_{R}^{c}}d_{R}S_{1}^{*}
  + {y^{'}_{R}}\overline{d}{_{R}^{c}}N_{R}S_{1}+ {{y^{''}_{R}}\overline{\mathbb{D}}{_{R}^{c}}N_{R}S_{1}}\nonumber\\
  &&+{y^{'''}_{R}}\overline{u}{_{R}^{c}}e_{R}S_{1}+{y_{LR}}\overline{Q}_{L}\widetilde{\phi} \mathbb{D}_{R}+h.c.
  \label{Lagrangian}
 \end{eqnarray}

  
 
 SM Lagrangian and the Lagrangian obtained in (\ref{Lagrangian}) should undoubtedly be interpreted in the perspective of effective field theory. In the framework of effective operators, the $R_{D^{(\ast)}}$ anomalies has been addressed to a model-independent study that takes into account the impacts of the renormalization-group(RG) evolution \cite{Sakaki:2013bfa,Freytsis:2015qca}. More specifically in \cite{Freytsis:2015qca} experimental results are extremely well fit, and the relationship between leptoquark mass and coupling is established.
The interactions parameters in Lagrangian (\ref{Lagrangian}) can explain new physics contributions to the flavor anomalies by constraining the LQ coupling and masses.
 A comprehensive analysis of the significant contribution of new physics to the aforementioned anomalies is skipped here. The interested reader may go through recent works \cite{Dorsner:2016wpm,Parashar:2022wrd} and the references therein.

From a phenomenological perspective, the TeV-scale $S_1$ scalar leptoquark in a GUT framework is extremely exciting because its existence can be established at the LHC \cite{Angelescu:2021lln,Angelescu:2018tyl}. Pair produced leptoquarks have been looked for by ATLAS and CMS in a variety of final states. Exclusion constraints on $S_1$ have been established using various decay models by direct detection searches for scalar leptoquarks \cite{CMS:2018qqq,CMS:2018svy}.
Whereas the direct LHC bounds on this leptoquark are not particularly strict, as was emphasized in reference \cite{Mandal:2018kau}, the most recent LHC data from the $pp\rightarrow \tau\tau/\tau \nu$ channels have actually imposed restrictions on the parameter space of $S_1$ necessary to describe and understand $R_{D^{(\ast)}}$ anomalies.

Thus, the B-physics anomalies can be explained in presence of vector-like fermions and scalar leptoquark at the TeV scale.

{\bf 5. Conclusion:\,}
 
 We have investigated a non-supersymmetric $E_6$ GUT, with fermions taken from the fundamental representation $27$ and scalars from  fundamental $27$ as well as the adjoint representation $78$. With the minimal version and with no intermediate symmetry, the model  provides successful unification of the fundamental forces of the Standard Model. With GUT threshold effects, it can simultaneously addresses some unsolved puzzles of SM in tune with the phenomenological constraints. The model with additional sterile neutrinos, can predict sub-eV scale neutrino at one-loop level via MRIS mechanism. 

 Another interesting feature of the $E_6$ GUT in this minimal theory with scalar leptoquark and vector-like fermions at the TeV scale, is that one can appreciate the muon $g-2$ anomaly reported by the Fermilab $g-2$ experiment \cite{Muong-2:2006rrc,Blum:2013xva,ParticleDataGroup:2020ssz,Muong-2:2021ojo} and its suitable explanation. The new contributions to the muon magnetic dipole moment $a_{\mu}$ \cite{Jegerlehner:2009ry,Lindner:2016bgg,Patra:2016shz,Majumdar:2020xws} can arise from the interactions of vector-like fermions $\mathbb{D}_{L,R} (3,1,-1/3)$, $\Psi_{L,R} (1,2,-1/2)$ and a scalar leptoquark $S_{1}(\overline{3},1,1/3)$ with the SM fields and muons. Thus the proposed $E_6$ Grand Unified Theory (GUT),   in it’s simple form and with no intermediate symmetry can be an all-in-one model for explanation of neutrino mass, B-physics anomalies and muon $g-2$ anomaly.

{\bf Acknowledgement:}
Chandini Dash would like to thank Department 
of Science and Technology, Govt. of India for INSPIRE Fellowship/2015/IF150787 in support of her research work.

\bibliographystyle{utcaps_mod}
\bibliography{E6_SM}

\end{document}